\documentclass[english]{article}

\usepackage{scalerel}
\usepackage{tikz}
\usetikzlibrary{svg.path}

\definecolor{orcidlogocol}{HTML}{A6CE39}
\tikzset{
  orcidlogo/.pic={
    \fill[orcidlogocol] svg{M256,128c0,70.7-57.3,128-128,128C57.3,256,0,198.7,0,128C0,57.3,57.3,0,128,0C198.7,0,256,57.3,256,128z};
    \fill[white] svg{M86.3,186.2H70.9V79.1h15.4v48.4V186.2z}
                 svg{M108.9,79.1h41.6c39.6,0,57,28.3,57,53.6c0,27.5-21.5,53.6-56.8,53.6h-41.8V79.1z M124.3,172.4h24.5c34.9,0,42.9-26.5,42.9-39.7c0-21.5-13.7-39.7-43.7-39.7h-23.7V172.4z}
                 svg{M88.7,56.8c0,5.5-4.5,10.1-10.1,10.1c-5.6,0-10.1-4.6-10.1-10.1c0-5.6,4.5-10.1,10.1-10.1C84.2,46.7,88.7,51.3,88.7,56.8z};
  }
}

\newcommand\orcidicon[1]{\href{https://orcid.org/#1}{\mbox{\scalerel*{
\begin{tikzpicture}[yscale=-1,transform shape]
\pic{orcidlogo};
\end{tikzpicture}
}{|}}}}

\usepackage[utf8]{inputenc}
\usepackage[T1]{fontenc}
\usepackage{babel}
\usepackage{amsmath}
\usepackage{graphicx}
\usepackage{fancyhdr}
\usepackage{siunitx}
\usepackage[export]{adjustbox}
\usepackage{array}
\usepackage{subcaption}
\usepackage{enumitem}
\usepackage{multirow}
\usepackage{booktabs}
\usepackage{siunitx}
\usepackage{bm}
\usepackage{makecell}
\usepackage{pbox}
\usepackage{xcolor}
\usepackage{hyperref}

\usepackage[style=nature]{biblatex}
\newcommand{\scidatalogo}{\includegraphics[height=36pt]{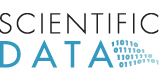}}
\newcommand{\overleaflogo}{\includegraphics[height=36pt]{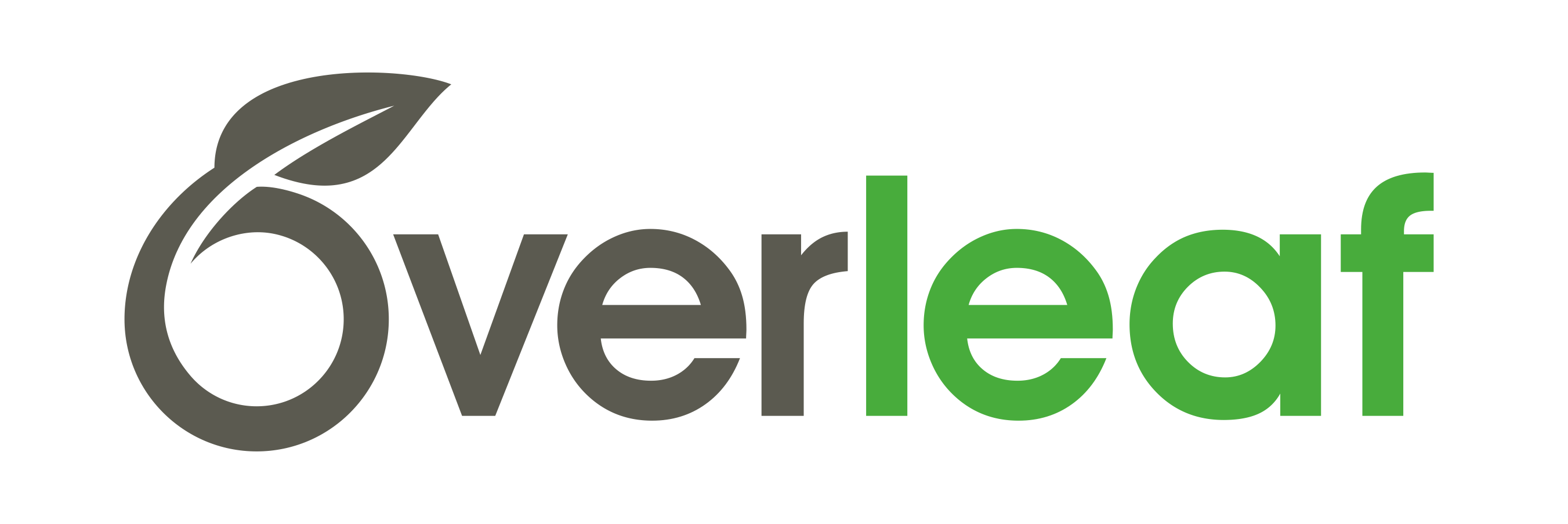}}
\begin{document}

\title{OutFin, a multi-device and multi-modal dataset for outdoor localization based on the fingerprinting approach}

 \author{Fahad Alhomayani\textsuperscript{1} \orcidicon{0000-0002-4914-8722} and Mohammad H. Mahoor\textsuperscript{1{*}}}

\maketitle
\thispagestyle{fancy}

 1. Department of Electrical and Computer Engineering, Ritchie School of Engineering and Computer Science, University of Denver, Denver, CO, 80208 USA {*}corresponding author: Dr. Mohammad H. Mahoor (mmahoor@du.edu)
\begin{abstract}
In recent years, fingerprint-based positioning has gained researchers’ attention since it is a promising alternative to the Global Navigation Satellite System and cellular network-based localization in urban areas. Despite this, the lack of publicly available datasets that researchers can use to develop, evaluate, and compare fingerprint-based positioning solutions constitutes a high entry barrier for studies. As an effort to overcome this barrier and foster new research efforts, this paper presents OutFin, a novel dataset of outdoor location fingerprints that were collected using two different smartphones. OutFin is comprised of diverse data types such as WiFi, Bluetooth, and cellular signal strengths, in addition to measurements from various sensors including the magnetometer, accelerometer, gyroscope, barometer, and ambient light sensor. The collection area spanned four dispersed sites with a total of \si{122} reference points. Each site is different in terms of its visibility to the Global Navigation Satellite System and reference points' number, arrangement, and spacing. Before OutFin was made available to the public, several experiments were conducted to validate its technical quality. 
\end{abstract}

\section*{Background \& Summary}

Location-Based Services (LBS) has become a multibillion-dollar industry that is expected to continue to steadily grow over the upcoming years \cite{alliedmarketresearch}. Some of these services include location-based marketing \cite{hopkins2012go}, authentication \cite{hammad2017location}, gaming \cite{leorke2014location}, and social networking \cite{zheng2011location}, among others. A key enabling technology at the heart of such services is positioning \cite{doi:10.1080/17489725.2018.1508763}. However, the de facto standard for positioning, the Global Navigation Satellite System (GNSS), has two major issues that limit the use of LBS. First, the availability and accuracy of GNSS are severely degraded in urban areas due to shadowing and multipath effects \cite{ranacher2016gps}. Second, GNSS chipsets are notorious for being power-hungry, which is problematic for power-constrained devices such as smartphones and smartwatches \cite{carroll2010analysis}. A more energy-efficient approach for positioning is achieved using cellular networks. Yet, the offered accuracy, which is in the order of tens \cite{10.1145/1999995.2000024} to hundreds \cite{zandbergen2009accuracy} of meters, fails to satisfy the accuracy requirements imposed by many services and applications. 

\begin{figure}[!b]
\centering
\includegraphics[width=\textwidth]{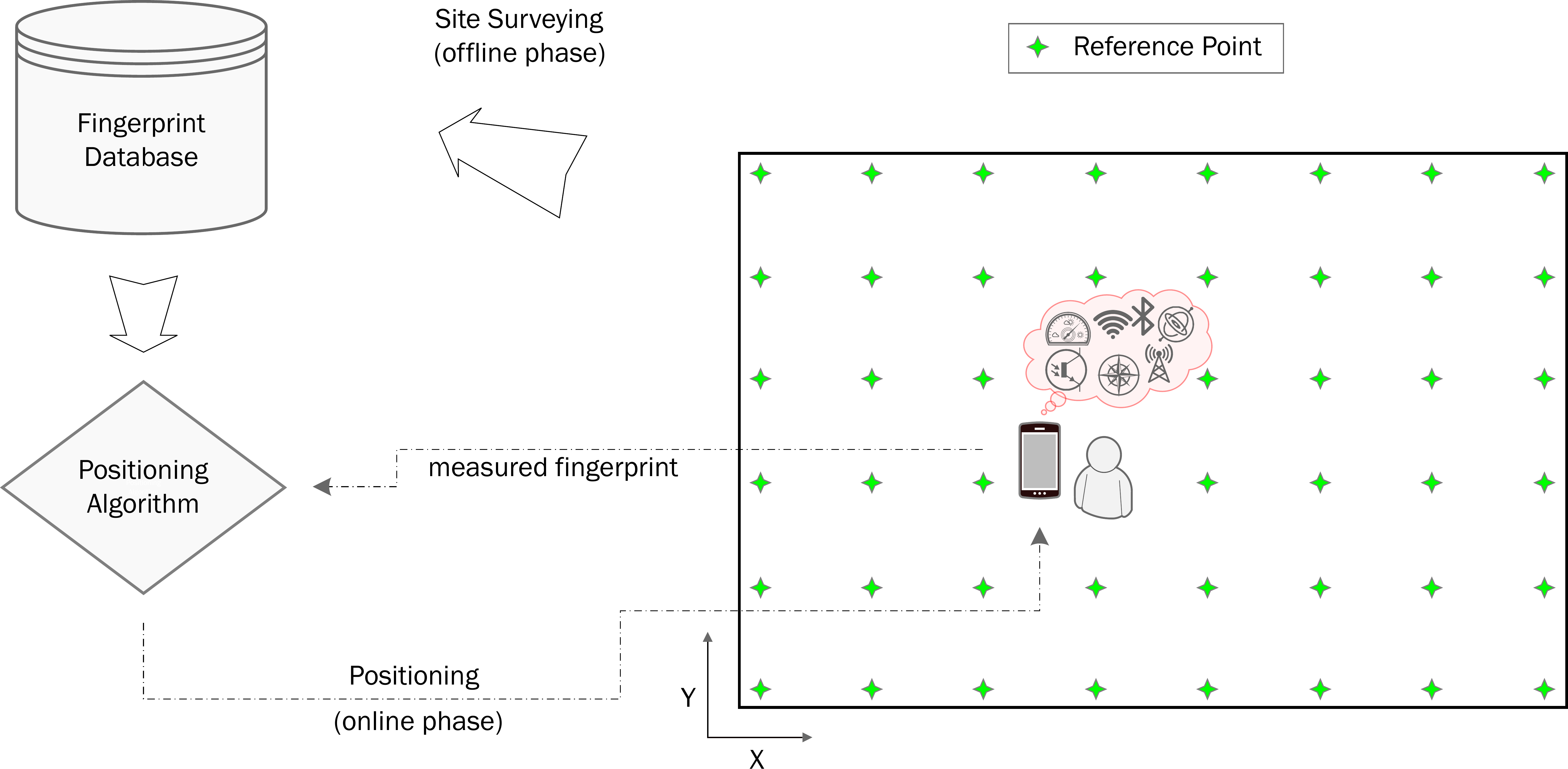}
\caption{A graphical representation of the fingerprinting approach for positioning.}
\label{fingerprint_process}
\end{figure}

Recently, in an attempt to devise positioning solutions that can yield better performance, researchers have turned their attention to \textit{fingerprinting}, a positioning technique that has achieved great success in the indoor positioning domain, a domain where GNSS signals are generally unavailable \cite{vo2015survey}. Fingerprinting is used to identify spatial locations based on location-dependent measurable features (location fingerprints). These fingerprints can be of different types such as WiFi fingerprints \cite{bahl2000radar}, Bluetooth fingerprints \cite{7103024}, cellular fingerprints \cite{8570849}, and magnetic field fingerprints \cite{8626558}. From an implementation perspective, the fingerprinting approach is a two-phase process that consists of an \textit{offline phase} and an \textit{online phase}. During the offline phase, \textit{site surveying} is performed by sampling fingerprints of an area of interest at predefined \textit{reference points} (RPs). Fingerprints are often sampled using a smartphone or a dedicated data acquisition platform. Fingerprints, along with the coordinates at which they were sampled, are stored in a database. The data is then used to train a machine learning algorithm to learn a function that best maps sampled fingerprints to their ground truth coordinates. Afterward, the learned function is utilized during the online phase to infer a user’s coordinates given the fingerprints measured at the user’s location. The process of fingerprinting is visually depicted in Fig. \ref{fingerprint_process}.

Despite its low complexity and ability to produce accurate location estimates, the main drawback of fingerprinting is the laborious and time-consuming site surveying task. This drawback has led many studies to resort to either simulated \cite{luo2016deep} or crowdsourced data \cite{wang2016indoor}, where the former never fully reflects the real world and the latter may suffer from integrity and consistency problems. The proposal of OutFin aims at addressing these drawbacks by making real-world measurements and reliable ground truth coordinates publicly available.
Table \ref{table1} summarizes the main aspects of publicly available fingerprinting datasets published since 2014. Compared to these datasets, OutFin combines several features that place it in a unique position:
\begin{itemize}
  \item To the best of our knowledge, OutFin is the first multi-modal, outdoor fingerprints dataset to be publicly available.
  \item The data was collected using two contemporary smartphones rather than outdated smartphones or custom-built platforms.
  \item The data was collected at highly granular RPs with 61 to 183 centimeters (cm) spacing.
  \item OutFin not only provides location fingerprints, but it also provides information about the devices that generated them (e.g., the service set identifier of an access point, the communication protocol of a Bluetooth device, and the number of neighboring cells of a serving cell).
  \item OutFin is accompanied by an interactive map that provides various information about the collection environment, such as RP coordinates (both ground truth and Global Positioning System (GPS) estimates) and building ground elevations and heights.
\end{itemize}

\begin{table}[!b]
\centering
\begin{adjustbox}{width=1.2\textwidth,center=\textwidth}
\begin{tabular}{@{}m{3cm}m{1cm}m{1.6cm}m{4cm}m{3cm}m{3cm}m{2.2cm}m{2cm}@{}} 
\toprule
\bfseries Dataset & \bfseries Year	& \bfseries Category & \bfseries Environment & \bfseries Data type(s) & \bfseries Device type(s) & \bfseries \# of samples & \bfseries Granularity\\  
\midrule
UJIIndoorLoc \cite{7275492} & 2014 & Indoor & Three university buildings & WiFi & Smartphone, Tablet & Tens of thousands & Medium\\ 
\midrule
UJIIndoorLoc-Mag \cite{7346763} & 2015 & Indoor & A research lab & sensor & Smartphone & Tens of thousands & Medium\\
\midrule
Dataset described in \cite{7743678} & 2016 & Indoor & A research facility & WiFi, sensor & Smartphone, Smartwatch & Tens of thousands & High\\
\midrule
Dataset described in \cite{7477348} & 2016 & Indoor & A university building & WiFi, Bluetooth, sensor & Smartphone & Thousands & High\\
\midrule
PerfLoc \cite{7794983} & 2016 & Indoor & An office building, two industrial warehouses, and a subterranean structure & WiFi, cellular, sensor & Smartphone & Millions & Medium\\
\midrule
AmbiLoc \cite{ popleteev2017ambiloc} & 2017 & Indoor & An apartment and two university buildings & TV, FM, cellular & Dedicated data acquisition platform & Thousands & Medium\\
\midrule
MagPIE \cite{8115961} & 2017 & Indoor & Three university buildings & sensor & Smartphone & Hundreds of thousands & High\\
\midrule
Dataset described in \cite{mendoza2018long} & 2018 & Indoor & A university library & WiFi & Smartphone & Hundreds of thousands & High\\
\midrule
Dataset described in \cite{byrne2018residential} & 2018 & Indoor & Four residential homes & Bluetooth, sensor & Dedicated data acquisition platform & Hundreds of thousands & High\\
\midrule
Dataset described in \cite{7945258} & 2018 & Indoor & A university library & Bluetooth & Smartphone & Thousands & Medium\\
\midrule
Dataset described in \cite{baronti2018indoor} & 2018 & Indoor & A research facility & Bluetooth & Smartphone, Dedicated data acquisition platform & Millions & High\\
\midrule
Dataset described in \cite{aernouts2018sigfox} & 2018 & Outdoor & A large-scale urban area and a large-scale rural area & Sigfox, LoRaWAN & Dedicated data acquisition platform & Hundreds of thousands & Low\\
\midrule
Dataset described in \cite{mendoza2019ble} & 2019 & Indoor & Two university buildings & Bluetooth & Smartphone & Thousands & High\\
\midrule
Dataset described in \cite{10.1007/978-3-030-30278-8_14} & 2019 & Indoor, Outdoor & Worldwide & Cellular & Smartphone & Millions & Low\\
\midrule
\textbf{OutFin} \cite{OutFinData} & \textbf{2020} & \textbf{Outdoor} & \textbf{A university campus} & \textbf{WiFi, Bluetooth, cellular, sensor} & \textbf{Smartphone} & \textbf{Hundreds of thousands} & \textbf{High}\\
\bottomrule
\end{tabular}
\end{adjustbox}
\caption{A comparison of the main aspects of publicly available fingerprinting datasets published since 2014. \textbf{Dataset}: the name of the dataset (if indicated) and a reference to its description. \textbf{Year}: the year the dataset was made available. \textbf{Category}: indicates whether the data was collected indoors or outdoors. \textbf{Environment}: a brief description of the collection environment. \textbf{Data type(s)}: the type(s) of data that was collected. \textbf{Device type(s)}: the type(s) of devices used to collect the data. \textbf{\# of samples}: the highest place value of the number of samples in the dataset. \textbf{Granularity}: a descriptor indicating how close the RPs were to each other; High: indicates a spacing of fewer than 2 meters, Medium: indicates a spacing between 2 and 8 meters, and Low: indicates a spacing of greater than 8 meters.}
\label{table1}
\end{table}

In addition to facilitating the research and development of outdoor positioning solutions that are based on the fingerprinting approach, OutFin might spur innovation in other research realms, including but not limited to: machine learning \cite{vepakomma2018}, Bayesian optimization \cite{NIPS2018_7472}, simultaneous localization and mapping \cite{8279260}, and map-matching \cite{8559918}.

\section*{Methods}
\subsection*{Data acquisition platform}
OutFin was created using two smartphones for data acquisition: Samsung’s Galaxy S10+ (Phone 1) and Google’s Pixel 4 (Phone 2). The former was released in the U.S. market on March 8, 2019, while the latter was released on October 24, 2019. Both smartphones ran on Android 10, released on September 3, 2019. The motivation behind choosing Android-powered smartphones was twofold. First, Android provides application programming interfaces (APIs) that allow for acquiring raw data at the hardware level. Second, Android-powered smartphones account for over \SI{74}{\percent} of the market share worldwide \cite{statcounter}. The two smartphones were attached to a tripod head using a dual mount that horizontally separated them by \SI{10}{\cm} (see Fig. \ref{sites} (Site 1)). Both smartphones were in portrait mode. The tripod kept them at a fixed height of \SI{132}{\cm}. The tripod head was adjusted to tilt the smartphones at a $\sim$40 degree (\si{\degree}) angle to the vertical plane. The same set of third-party apps used for data collection were installed on both smartphones. These apps, which can be downloaded from the Google Play Store, included: WiFi Analyzer Pro (App 1) \cite{WiFiAnalyzer}, Bluetooth Scanner Extreme Edition (App 2) \cite{BluetoothScanner}, NetMonitor Pro (App 3) \cite{NetMonitorPro}, and Physics Toolbox Sensor Suite Pro (App 4) \cite{PhysicsToolbox}. The apps allowed for conveniently collecting and exporting WiFi, Bluetooth, cellular, and sensor data, respectively.

\subsection*{Data collection environment}
Data collection was performed at the University of Denver’s campus where four separate sites were considered. The motivation behind collecting data at separate sites was to offer diversity. For instance, each site is different in terms of its reference points’ number, arrangement, and spacing. Also, due to different ground elevations and heights of surrounding buildings, each site has different visibility to the GNSS. This is reflected by GPS errors produced at a given site. The mean GPS error was 12.1 meters (\si{\meter}), 11.4 m, 4.3 m, and 12.7 m for the first, second, third, and fourth site, respectively. GPS estimates are provided in OutFin to help researches compare their system’s performance to that obtained by GPS. A description of the data collection sites is provided below:

\begin{figure}[!b]
\centering
\begin{adjustbox}{width=1.3\textwidth,center=\textwidth}
\includegraphics{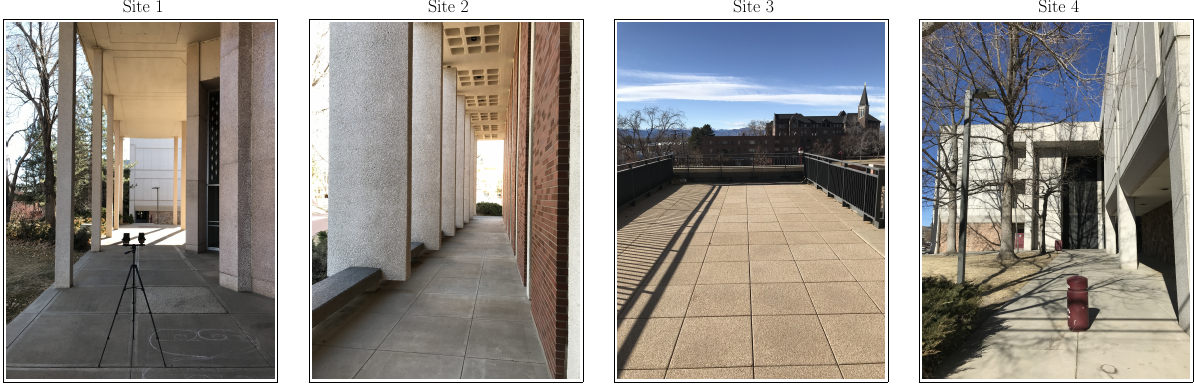}
\end{adjustbox}
\caption{Pictures of the four sites where data was  collected}
\label{sites}
\end{figure}

\begin{figure}[!b]
\centering
\begin{adjustbox}{width=1.6\textwidth,center=\textwidth}
\includegraphics{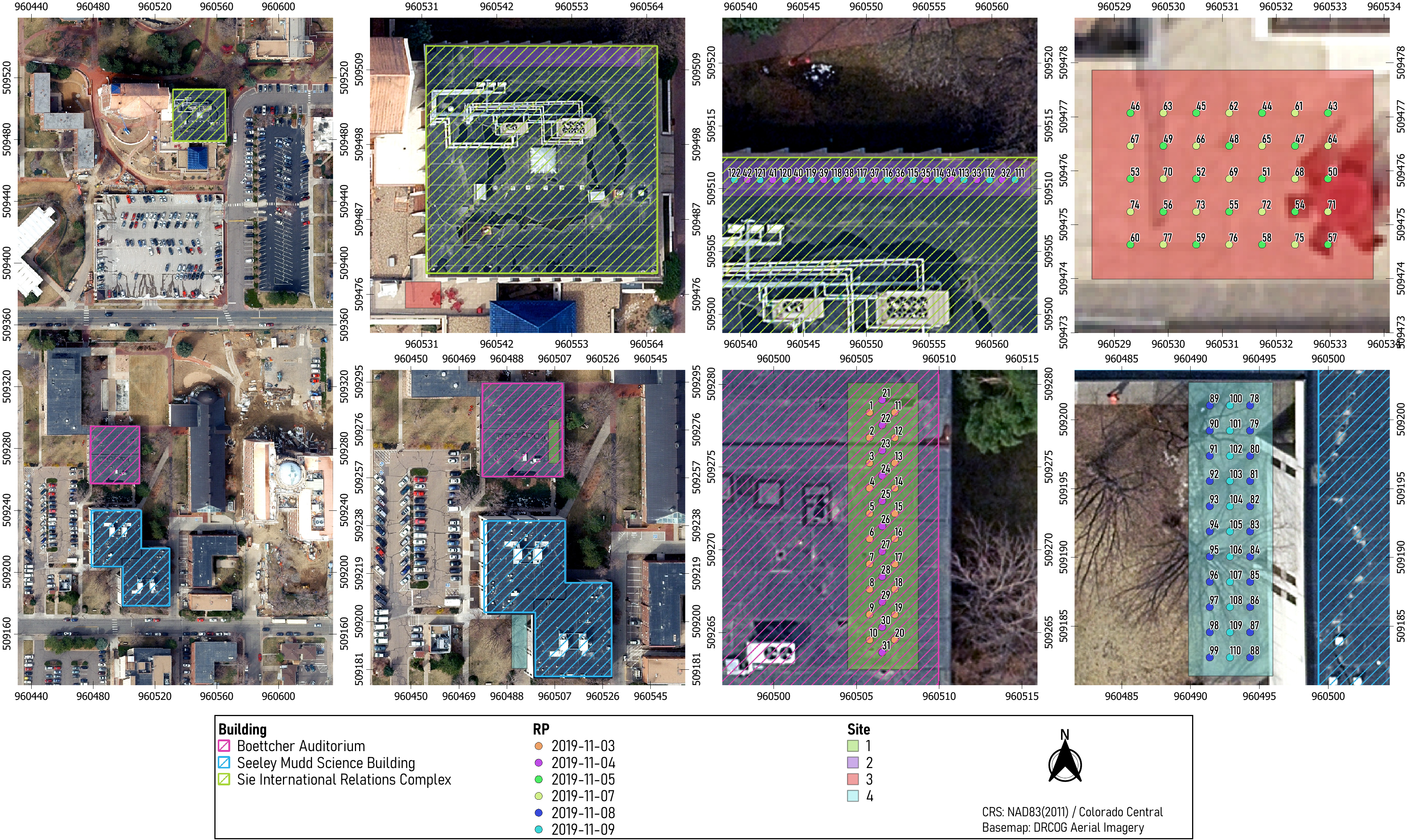}
\end{adjustbox}
\caption{An aerial map of the collection environment showing the four collection sites and the 122 RPs. RPs are color-coded according to the date of collection.}
\label{Outdoor_Map}
\end{figure}

\begin{enumerate}[label=Site \arabic{enumi}:,ref=Site \arabic{enumi}, leftmargin=*]
\item Site 1 represents a portion of a covered sidewalk next to the east side of the 11.8  \si{\meter} high Boettcher Auditorium (see Fig. \ref{sites}). Site 1 contained 31 RPs arranged in three north-to-south lines (see Fig. \ref{Outdoor_Map}). The spacing between RPs in each line was fixed at \SI{152.5}{\cm} and the distance between lines was fixed at \SI{76.25}{\cm}. 

\item Site 2 is $\sim$\SI{245}{\meter} north of Site 1 and represents a portion of a covered sidewalk next to the north side of the \SI{11.5}{\meter} high Sie International Relations Complex (see Fig. \ref{sites}). Site 2 contained 23 RPs arranged in a single east-to-west line (see Fig. \ref{Outdoor_Map}). The spacing between RPs was fixed at \SI{101.5}{\cm}. 

\item Site 3 is $\sim$\SI{40}{\meter} south of Site 2 and represents a portion of an open terrace next to the south side of the Sie International Relations Complex (see Fig. \ref{sites}). Site 3 contains 35 RPs arranged in a seven-column and five-row grid (see Fig. \ref{Outdoor_Map}). The spacing between column RPs and row RPs were fixed at \SI{61}{\cm}. 

\item Site 4 is $\sim$\SI{288}{\meter} south of Site 3 and represents a portion of an open sidewalk by the south and west sides of the \SI{13.4}{\meter} high Seeley Mudd Science Building (see Fig. \ref{sites}). Site 4 contains 33 RPs arranged in a three-column and eleven-row grid (see Fig. \ref{Outdoor_Map}). The spacing between column RPs was fixed at \SI{183}{\cm}, while the spacing between row RPs was fixed at \SI{146.5}{\cm}.
\end{enumerate}

Each RP is uniquely identified by an integer (an ID number) that symbolizes its order in the collection campaign. For example, data collection started with RP 1 on November 3, 2019, and ended with RP 122 on November 9, 2019. The ground truth locations of RPs belonging to a site are expressed with respect to a local frame of reference. Additionally, the easting and northing (X,Y) coordinates of all RPs were provided with respect to a global coordinate system (i.e., NAD83(2011)/Colorado Central). This was accomplished with help from the university’s Department of Geography \& the Environment and by using a geographic information system software \cite{QGIS}.

\subsection*{Procedure}
Data collection spanned six days (3--5/11/2019 and 7--9/11/2019) and involved four sites with a total of 122 RPs. \textcolor{red}{Due to the fact that rain could severely affect wireless signal measurements, we did not collect any data on rainy days.} The RPs surveyed each day are indicated in Fig. \ref{Outdoor_Map}. The sequence of steps performed during a day of data collection are described below:

\begin{enumerate}[label=Step \arabic{enumi}:,ref=Step \arabic{enumi}, leftmargin=*]
\item Before mounting the smartphones to the tripod, App 4 was launched to collect magnetic field measurements by rotating the smartphones around their X, Y, and Z axes multiple times (see Fig. \ref{phone}). This process was performed for at least two minutes at a sampling rate of 1 Hertz (\si{\hertz}). The resultant data was exported as a comma-separated values (CSV) file, named with the smartphone’s name and date (e.g., \texttt{Phone1\_051119.csv}). Such data can be used to offset the hard-iron distortion caused by placing the smartphones close to each other. After this process, the smartphones were mounted to the tripod and placed at the RP where data was to be collected.

\begin{figure}[!b]
\centering
\begin{adjustbox}{width=0.4\textwidth,center=\textwidth}
\includegraphics{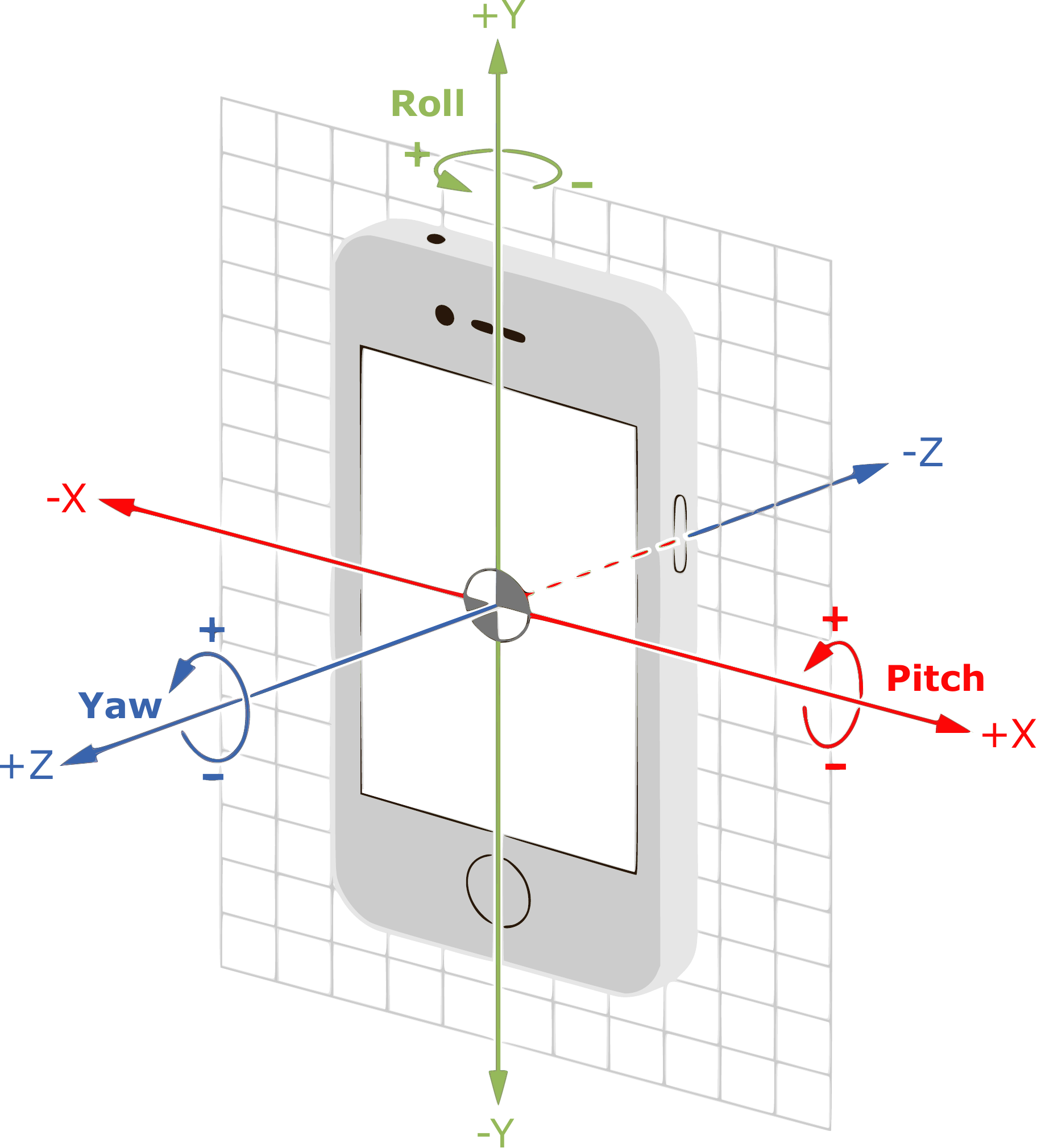}
\end{adjustbox}
\caption{Illustration of the X, Y, and Z axes relative to a typical smartphone. Figure reproduced from \cite{ReferenceFrame}.}
\label{phone}
\end{figure}

\item App 1 was launched to collect WiFi data, ensuring that at least two WiFi scans were performed along the four cardinal directions by routing the tripod head counterclockwise, $\sim$\SI{90}{\degree} at a time. A WiFi scan recorded the received signal strength (RSS) from all access points (APs) in range in addition to information about the APs themselves. Android only supports passive scanning, and the duration of a scan varies depending on the smartphone’s WiFi hardware and firmware. However, Google recently released a restriction that limits the frequency of scans that an app can perform to only four times in a 2-minute period \cite{android_WiFi}. This restriction applies to Android 9 and higher. The app reported scan results approximately every 30 seconds for Phone 1 and every 25 seconds for Phone 2. For Site 1 and 4's RPs, data collection started facing south and ended facing west. For Site 2 and 3's RPs, data collection started facing west and ended facing north. Collecting data along four directions mitigates the shadowing effect caused by the body of the data collector who is constantly facing the smartphone screens. Scan outcomes were exported as a CSV file, named with the smartphone’s model as a prefix and the RP’s ID as a suffix (e.g., \texttt{Phone2\_WiFi\_73.csv}).

\item App 2 was launched to collect Bluetooth data. Android allows active Bluetooth scanning; thus, scans can be triggered by a user-level app. A Bluetooth scan involves an inquiry scan of approximately 12 seconds, followed by a page scan for each discovered device to retrieve its information and the RSS \cite{android_bluetooth}. The duration of a scan, for both smartphones, took anywhere between 15 and 30 seconds, primarily depending on the number of discoverable devices in the area. As in Step 2, the shadowing effect was accounted for by performing two scans along each cardinal direction. Scan results were exported as a CSV file with a naming convention like that described in Step 2 (e.g., \texttt{Phone1\_Bluetooth\_29.csv}). 

\item App 3 was launched to collect cellular data. A smartphone’s cellular modem constantly scans the cellular network for cell selection/reselection and handover purposes. Android provides APIs to extract information associated with scans such as Reference Signal Received Power (RSRP) and cell identity information \cite{android_telephony}. The sampling frequency can be set manually and was fixed to \SI{1}{\hertz}. As noted in Step 2, the shadowing effect was accounted for by collecting at least fifteen samples along each cardinal direction. Collected data was exported as a CSV file with a naming convention like that described previously (e.g., \texttt{Phone2\_Cellular\_14.csv}). Moreover, App 3 allowed for collecting GPS data as part of the data record. The GPS readings corresponding to RPs belonging to the same site were extracted and stored under a CSV file named with the site's name as a prefix and the smartphone’s model and app name as a suffix (e.g., \texttt{Site1\_GPS\_Phone1\_App3.csv}). 
\item App 4 was launched to collect sensor data. A smartphone's built-in sensors can be classified as either hardware-based, such as the magnetometer and gyroscope, or software-based, such as the gravity and linear acceleration sensors. Android provides APIs for accessing and acquiring raw sensor data at defined rates \cite{android_sensor}. The sampling frequency was set to \SI{1}{\hertz}. Although sensor measurements are not subject to the shadowing effect, data was collected along the four cardinal directions to both conform with the survey pattern established above and diversify the dataset since magnetic field strength can vary greatly even within a small area (in the orders of a few centimeters or less) \cite{6418864}. At least fifteen samples were collected along each direction, following the same directions described in Step 2. Sensor data was exported as a CSV file with a naming convention like that described previously (e.g., \texttt{Phone1\_Sensors\_58.csv}). App 4 also allowed for collecting GPS data as part of the data record. As in Step 4, the GPS readings corresponding to RPs belonging to the same site were extracted and stored under a CSV file with a naming convention like
that described in Step 4 (e.g., \texttt{Site3\_GPS\_Phone2\_App4.csv}).
\item The tripod was moved to the next RP and Steps 2--5 were repeated. This process continued until all RPs designated for a given day were surveyed.
\end{enumerate}

\section*{Data Records}

On April 2, 2020, the OutFin dataset was made publicly available on figshare \cite{OutFinData}. Fig. \ref{directory} shows the dataset’s file structure and presents an overview of all CSV file types, their field labels, and a data record example. A description of the CSV file types and their field labels is provided below:

\begin{figure}[!b]
\centering
\begin{adjustbox}{width=1.5\textwidth,center=\textwidth}
\includegraphics{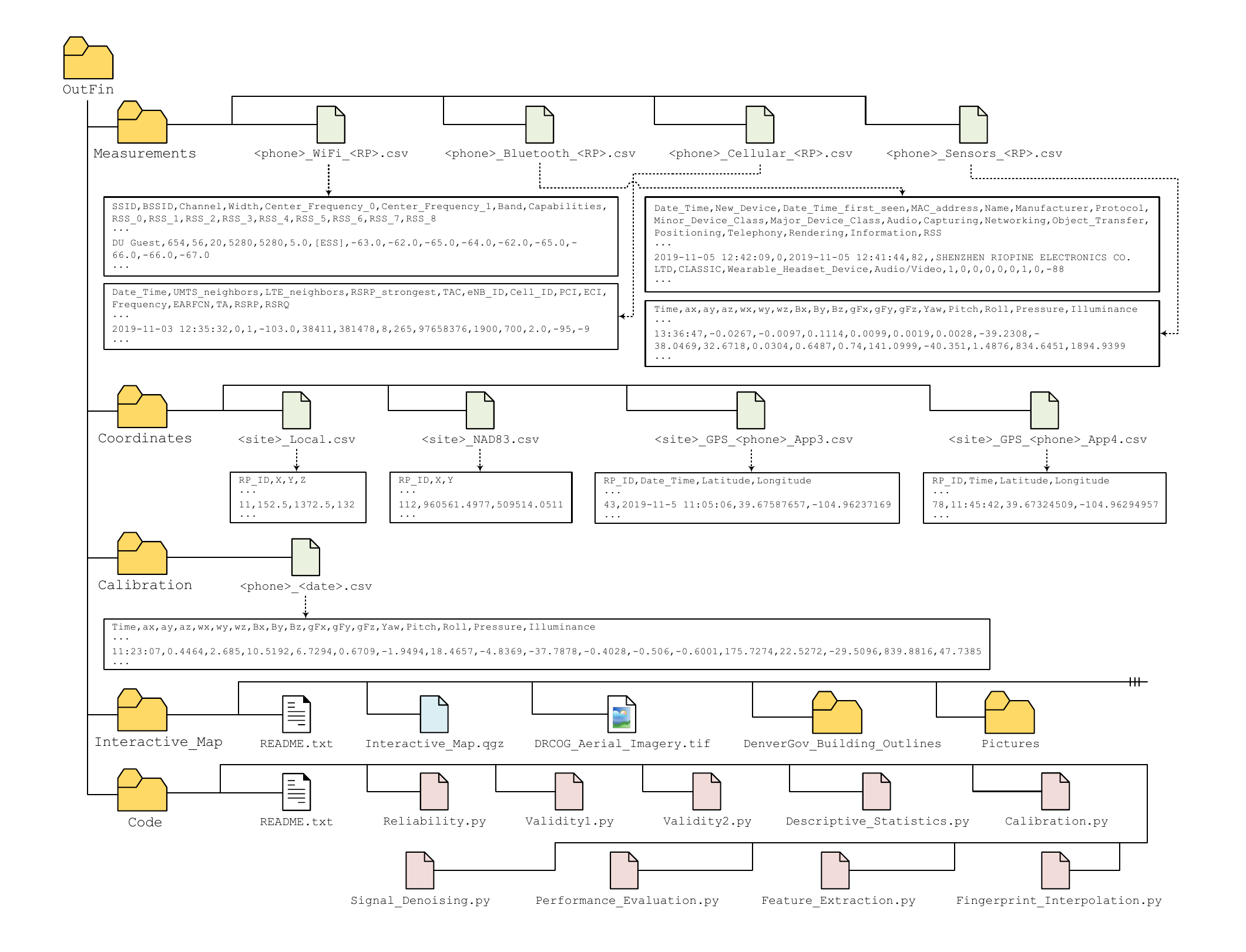}
\end{adjustbox}
\caption{Directory tree of the OutFin dataset along with CSV file types and example data records. \texttt{<phone>} $\in$ \{\texttt{Phone1,Phone2}\}, \texttt{<RP>} $\in$ \{\texttt{1,2,...,122}\}, \texttt{<site>} $\in$ \{\texttt{Site1,Site2,Site3,Site4}\}, and \texttt{<date>} $\in$ \{\texttt{031119,041119,051119,071119,081119,091119}\}.} 
\label{directory}
\end{figure}

\begin{itemize}
\item [I.] \texttt{<phone>\_WiFi\_<RP>.csv} contains WiFi data collected by a smartphone via App 1:
\begin{itemize}
\item [1.] \texttt{SSID}: The Service Set IDentifier (i.e., the AP’s network name).
\item [2.] \texttt{BSSID}: The Basic Service Set IDentifier (i.e., the AP’s media access control address (MAC address)) encoded as an integer. 
\item [3.] \texttt{Channel}: The channel number that the AP uses for communication.
\item [4.] \texttt{Width}: The bandwidth of the channel in megahertz (MHz); can be \texttt{20}, \texttt{40}, or \texttt{80} MHz.
\item [5.] \texttt{Center\_Frequency\_0}: The center frequency of the primary channel in MHz. 
\item [6.] \texttt{Center\_Frequency\_1}: The center frequency of the 40 or 80 MHz-wide channel in MHz. If a 20-MHz channel is used, then \texttt{Center\_Frequency\_1} $\equiv$ \texttt{Center\_Frequency\_0}.
\item [7.] \texttt{Band}: The AP’s frequency band in gigahertz (GHz); can be either \texttt{2.4} or \texttt{5} GHz.
\item [8.] \texttt{Capabilities}: Describes the authentication, key management, and encryption schemes supported by the AP.
\item [9--17.] \texttt{RSS\_0--RSS\_8}: The Received Signal Strengths in decibel-milliwatts (dBm), with respect to the back-to-back scans. 
\end{itemize}

\item [II.] \texttt{<phone>\_Bluetooth\_<RP>.csv} contains Bluetooth data collected by a smartphone via App 2:
\begin{itemize}
\item [1.] \texttt{Date\_Time}: The date and time the scan was triggered as \texttt{YYYY-MM-DD} and \texttt{hh:mm:ss}. Denver, Colorado is in the Mountain Time Zone, which is seven hours behind Coordinated Universal Time (UTC-07:00).
\item [2.] \texttt{New\_Device}: A binary flag that is set to 1 if the remote Bluetooth device is discovered for the first time at the current RP.
\item [3.] \texttt{Date\_Time\_first\_seen}: The date and time the device was first discovered at the current RP. The date and time formats are as described above.
\item [4.] \texttt{MAC\_address}: The device’s MAC address encoded as an integer.
\item [5.] \texttt{Name}: The device’s friendly name.
\item [6.] \texttt{Manufacturer}: The device’s manufacturer name.
\item [7.] \texttt{Protocol}: The Bluetooth protocol that the device uses for communication; can be \texttt{CLASSIC} (Basic Rate/Enhanced Data Rate (BR/EDR)), \texttt{BLE} (Bluetooth Low Energy), or \texttt{DUAL} (BR/EDR + BLE).
\item [8, 9.] \texttt{Minor\_Device\_Class}, \texttt{Major\_Device\_Class}: Indicates the device’s minor and major classes, respectively, as specified by the Bluetooth Special Interest Group (SIG) \cite{BluetoothSIG}. 
\item [10--17.] \texttt{Audio}, \texttt{Capturing}, \texttt{Networking}, \texttt{Object\_Transfer}, \texttt{Positioning}, \texttt{Telephony}, \texttt{Rendering}, \texttt{Information}: Binary flags that are set to 1 if the device is associated with any of the eight service classes specified by the Bluetooth SIG \cite{BluetoothSIG}.
\item [18.] \texttt{RSS}: The Received Signal Strength in dBm.
\end{itemize}

\item [III.] \texttt{<phone>\_Cellular\_<RP>.csv} contains cellular data collected by a smartphone via App 3. It should be noted that the entire collection environment was covered by Long-Term Evolution (LTE) cells. The Public Land Mobile Network (PLMN) identifier is 310410:
\begin{itemize}
\item [1.] \texttt{Date\_Time}: The date and time the sample was captured. The date and time formats are as described above.
\item [2.] \texttt{UMTS\_neighbors}: The number of neighboring Universal Mobile Telecommunications Service (UMTS) cells.
\item [3.] \texttt{LTE\_neighbors}: The number of neighboring LTE cells.
\item [4.] \texttt{RSRP\_strongest}: The Reference Signal Received Power, in dBm, corresponding to the strongest neighboring cell, which employs the same technology as the serving cell.
\item [5.] \texttt{TAC}: The Tracking Area Code, which uniquely defines a group of cells within a PLMN. 
\item [6.] \texttt{eNB\_ID}: The E-UTRAN (Evolved-UMTS Terrestrial Radio Access Network) NodeB IDentifier that is used to uniquely identify an eNB (i.e., a base station in LTE) within a PLMN.
\item [7.] \texttt{Cell\_ID}: The Cell IDentifier, which is an internal descriptor for a cell. It can take any value between \texttt{0} and \texttt{255}.
\item [8.] \texttt{PCI}: The Physical Cell Identifier that is used to indicate the physical layer identity of a cell. It can take any value between \texttt{0} and \texttt{503}.
\item [9.] \texttt{ECI}: The E-UTRAN Cell Identifier that is used to uniquely identify a cell within a PLMN. \texttt{ECI} = 256 $\times$ \texttt{eNB\_ID} + \texttt{Cell\_ID}.
\item [10.] \texttt{Frequency}: The downlink frequency band in MHz.
\item [11.] \texttt{EARFCN}: The downlink E-UTRAN Absolute Radio Frequency Channel Number.
\item [12.] \texttt{TA}: The Timing Advance value which ranges from \texttt{0} to \texttt{1282}. A change of 1 in TA corresponds to a 156\si{\meter} round-trip distance \cite{3GPP}. For example, if TA = 7, then the eNB is located within a 546\si{\meter} radius from the smartphone. 
\item [13.] \texttt{RSRP}: The Reference Signal Received Power in dBm. 	
\item [14.] \texttt{RSRQ}: The Reference Signal Received Quality in decibel (dB).
\end{itemize}

\item [IV.] \texttt{<phone>\_Sensors\_<RP>.csv} contains sensor data collected by a smartphone via App 4:
\begin{itemize}
\item [1.] \texttt{Time}: The time the sample was captured. The time format is as described above.
\item [2--4.] \texttt{ax}, \texttt{ay}, \texttt{az}: The linear acceleration, in meters per second squared (\si{m/s^2}), along the smartphone's X, Y, and Z axes, respectively.
\item [5--7.] \texttt{wx}, \texttt{wy}, \texttt{wz}: The angular velocity, in radian per second (\si{rad/s}), around the smartphone's X, Y, and Z axes, respectively.
\item [8--10.] \texttt{Bx}, \texttt{By}, \texttt{Bz}: The magnetic field strength, in microtesla (\si{\micro\tesla}), along the smartphone's X, Y, and Z axes, respectively.
\item [11--13.] \texttt{gFx}, \texttt{gFy}, \texttt{gFz}: The g-force measured as the ratio of normal force to gravitational force (FN/Fg), along the smartphone's X, Y, and Z axes, respectively.
\item [14--16.] \texttt{Yaw}, \texttt{Pitch}, \texttt{Roll}: The angle of rotation, in degrees (\si{\degree}), around the smartphone's X, Y, and Z axes, respectively.
\item [17.] \texttt{Pressure}: The atmospheric pressure in hectopascal (\si{\hecto\pascal}).
\item [18.] \texttt{Illuminance}: The illuminance in lux (\si{lx}).
\end{itemize}

\item [V.] \texttt{<site>\_Local.csv} contains the local coordinates of RPs belonging to a site. Each site has its own frame of reference and the origins are at RPs 10, 122, 60, and 99 for Sites 1, 2, 3, and 4, respectively.
\begin{itemize}
\item [1.] \texttt{RP\_ID}: The Reference Point IDentifier.
\item [2--4.] \texttt{X}, \texttt{Y}, \texttt{Z}: The X, Y, and Z coordinates of the RP in centimeters (cm).
\end{itemize}

\item [VI.] \texttt{<site>\_NAD83.csv} contains the global coordinates of RPs belonging to a site with respect to the NAD83(2011)/Colorado Central coordinate system.
\begin{itemize}
\item [1.] \texttt{RP\_ID}: The Reference Point IDentifier.
\item [2, 3.] \texttt{X}, \texttt{Y}: The X and Y coordinates of the RP in meters (m).
\end{itemize}

\item [VII.] \texttt{<site>\_GPS\_<phone>\_App3.csv} contains the GPS coordinates of RPs belonging to a site as computed by the smartphone's GPS chipset and reported by App 3.
\begin{itemize}
\item [1.] \texttt{RP\_ID}: The Reference Point IDentifier.
\item [2.] \texttt{Date\_Time}: The date and time the sample was captured. The date and time formats are as described above.
\item [3, 4.] \texttt{Latitude}, \texttt{Longitude}: The latitude and longitude coordinates of the RP.
\end{itemize}

\item [VIII.] \texttt{<site>\_GPS\_<phone>\_App4.csv} contains the GPS coordinates of RPs belonging to a site as computed by the smartphone's GPS chipset and reported by App 4.
\begin{itemize}
\item [1.] \texttt{RP\_ID}: The Reference Point IDentifier.
\item [2.] \texttt{Time}: The time the sample was captured. The time format is as described above.
\item [3, 4.] \texttt{Latitude}, \texttt{Longitude}: The latitude and longitude coordinates of the RP.
\end{itemize}

\item [IX.] \texttt{<phone>\_<date>.csv} contains sensors data collected by a smartphone via App 3 before the smartphone is mounted to the tripod. Field labels are identical to that described in IV (\texttt{<phone>\_Sensors\_<RP>.csv}).
\end{itemize}

\section*{Technical Validation}

The technical quality of the OutFin dataset was evaluated using experiments that consider two basic requirements that any high-quality dataset should satisfy, i.e., reliability and validity. Additionally, as a demonstration of the dataset’s potential for positioning applications, a number of practical usage examples are presented.

\textbf{Measurement Reliability}: A data acquisition platform is said to be reliable if it provides consistent measurements at different points in time. To this end, before the collection campaign, WiFi, Bluetooth, cellular, and sensor data was captured over three different days at the same location. Spearman's and Kendall's correlation coefficients were then used to quantify the degree of consistency between temporal measurements for a given phone. Table \ref{table2} shows Spearman's and Kendall's correlation coefficients for the two smartphones for all possible pairs of days. Given that correlation results are high (i.e., close to the maximum value of 1.0), it can be concluded that the dataset possesses a high degree of reliability. 

\begin{table}[!b]
\begin{adjustbox}{width=0.9\textwidth,center=\textwidth}
\resizebox{\textwidth}{!}{
\centering
	\begin{tabular}{@{}r*{6}{S[table-format=-3.4]}@{}}
	\toprule
	& \multicolumn{3}{c}{Phone 1} & \multicolumn{3}{c}{Phone 2} \\
	\cmidrule(lr){2-4} \cmidrule(lr){5-7}
	& {$\{day_1,day_2\}$} & {$\{day_2,day_3\}$} & {$\{day_1,day_3\}$} & {$\{day_1,day_2\}$} & {$\{day_2,day_3\}$} & {$\{day_1,day_3\}$} \\
	\cmidrule(lr){2-2} \cmidrule(lr){3-3} \cmidrule(lr){4-4} \cmidrule(lr){5-5} \cmidrule(lr){6-6} \cmidrule(lr){7-7} 
	WiFi\\
	\cmidrule(lr){1-1}
	\textit{Spearman's} $\rho$ & 0.960 & 0.949 & 0.946 & 0.952 & 0.968 & 0.936 \\ 
	\textit{Kendall's} $\tau$ & 0.837 & 0.826 & 0.815 & 0.828 & 0.877 & 0.796 \\ 
	\midrule
	Bluetooth\\
	\cmidrule(lr){1-1}
	\textit{Spearman's} $\rho$ & 0.575 & 0.736 & 0.700 & 0.716 & 0.889 & 0.790  \\    
	\textit{Kendall's} $\tau$ & 0.454 & 0.609 & 0.578 & 0.584 & 0.786 & 0.683 \\  
	\midrule
	Cellular\\ 
	\cmidrule(lr){1-1}
	\textit{Spearman's} $\rho$ & 0.964 & 0.964 & 1.0 &  0.964 & 0.964 & 1.0  \\  
	\textit{Kendall's} $\tau$ & 0.904 & 0.904 & 1.0 &  0.904 & 0.904 & 1.0  \\ 
	\midrule
	Sensors\\
	\cmidrule(lr){1-1}
	\textit{Spearman's} $\rho$ & 0.928 & 0.970 & 0.933 &  0.960 & 0.990 & 0.943  \\ 
	\textit{Kendall's} $\tau$ & 0.823 & 0.911 & 0.852 & 0.897 & 0.955 & 0.852 \\ 
	\bottomrule  
\end{tabular} 
} 
\end{adjustbox}
\caption{Results of the correlation analysis between the measurements obtained on three different days for Phone 1 and Phone 2. Spearman's $\rho$ varies between $-1$ and $+1$ with 0 implying no correlation, while values of $-1$ or $+1$ imply an exact monotonic relationship. Kendall’s $\tau$ varies between $-1$ and $+1$. Values close to $+1$ indicate strong agreement, while values close to $-1$ indicate strong disagreement. For WiFi, the results were generated using averaged RSS readings of fifty randomly selected APs that were observed over the three days. For Bluetooth, the results were generated using averaged RSS readings of fifteen randomly selected devices that were observed over the three days. The relatively lower correlation results obtained for Bluetooth is attributed to the fact that Bluetooth signals are more vulnerable to channel gain and fast fading than WiFi signals, causing measurements to fluctuate severely over time \cite{7103024}. For Cellular, the results were generated using averaged readings of UMTS neighbors, LTE neighbors, RSRP strongest, frequency, EARFCN, RSRP, and RSRQ from a cellular base station that a phone connected to over the three days. For Sensors, the results were generated using the averaged readings of linear acceleration, angular velocity, magnetic field strength, g-force, angle of rotation, atmospheric pressure, and illuminance. The \textit{p}-value of all results ranged between 0.0 and 0.02.} 
\label{table2}
\end{table}

\textbf{Measurement Validity}: A data acquisition platform is said to be valid if it accurately measures what it is intended to measure. In some cases, this requires the presence of theoretically-derived data to compare experimental data against. For example, WiFi RSS values can be computed using a path loss model. An input to the model is the distance between the transmitter and receiver. However, obtaining such inputs is not feasible since the exact location of all APs in the environment needs to be known. In the absence of theoretically-derived data, validity can be assessed by comparing data generated by different sources and checking for consistency. Accordingly, for a given day, Spearman's and Kendall's correlation coefficients were used to quantify the degree of consistency between the measurements obtained by the phones. The correlation results for the foregoing three days are shown in Table \ref{table3}. These results demonstrate high levels of consistency, which attests to the validity of the dataset.

\begin{table}[!b]
\begin{adjustbox}{width=0.5\textwidth,center=\textwidth}
\resizebox{\textwidth}{!}{
\centering
	\begin{tabular}{@{}r*{3}{S[table-format=-3.4]}@{}}
	\toprule
	& {$day_1$} & {$day_2$} &  {$day_3$}\\
	\cmidrule(lr){2-2} \cmidrule(lr){3-3} \cmidrule(lr){4-4}
	WiFi\\
	\cmidrule(lr){1-1}
	\textit{Spearman's} $\rho$ & 0.920 & 0.925 & 0.893\\  
	\textit{Kendall's} $\tau$ & 0.773 & 0.796 & 0.728 \\   
	\midrule
	Bluetooth\\
	\cmidrule(lr){1-1}
	\textit{Spearman's} $\rho$ & 0.763 & 0.706 & 0.843\\    
	\textit{Kendall's} $\tau$ & 0.657 & 0.535 & 0.703 \\  
	\midrule
	Cellular\\ 
	\cmidrule(lr){1-1}
	\textit{Spearman's} $\rho$ & 1.0 & 1.0 & 1.0 \\  
	\textit{Kendall's} $\tau$ & 1.0 & 1.0 & 1.0 \\      
	\midrule
	Sensors\\
	\cmidrule(lr){1-1}
	\textit{Spearman's} $\rho$ & 0.725 & 0.774 & 0.752 \\  
	\textit{Kendall's} $\tau$ & 0.617 & 0.720 & 0.676 \\       
	\bottomrule  
\end{tabular} 
}
\end{adjustbox}
\caption{Results of the correlation analysis between the measurements obtained from Phone 1 and Phone 2 for three different days. Spearman's $\rho$ varies between $-1$ and $+1$ with 0 implying no correlation, while values of $-1$ or $+1$ imply an exact monotonic relationship. Kendall’s $\tau$ varies between $-1$ and $+1$. Values close to $+1$ indicate strong agreement, while values close to $-1$ indicate strong disagreement. For WiFi, the results were generated using the averaged RSS readings of fifty randomly selected APs that were observed by both phones for a given day. For Bluetooth, the results were generated using the averaged RSS readings of fifteen randomly selected devices that were observed by both phones for a given day. For Cellular, the results were generated using averaged readings of UMTS neighbors, LTE neighbors, RSRP strongest, frequency, EARFCN, RSRP, and RSRQ of a cellular base station that both phones connected to for a given day. For Sensors, the results were generated using the averaged readings of linear acceleration, angular velocity, magnetic field strength, g-force, angle of rotation, atmospheric pressure, and illuminance for a given day. The \textit{p}-value of all results ranged between 0.0 and 0.01.} 
\label{table3}
\end{table} 

\begin{figure}[!h]
\centering
\begin{adjustbox}{width=1.06\textwidth,center=\textwidth}
\includegraphics{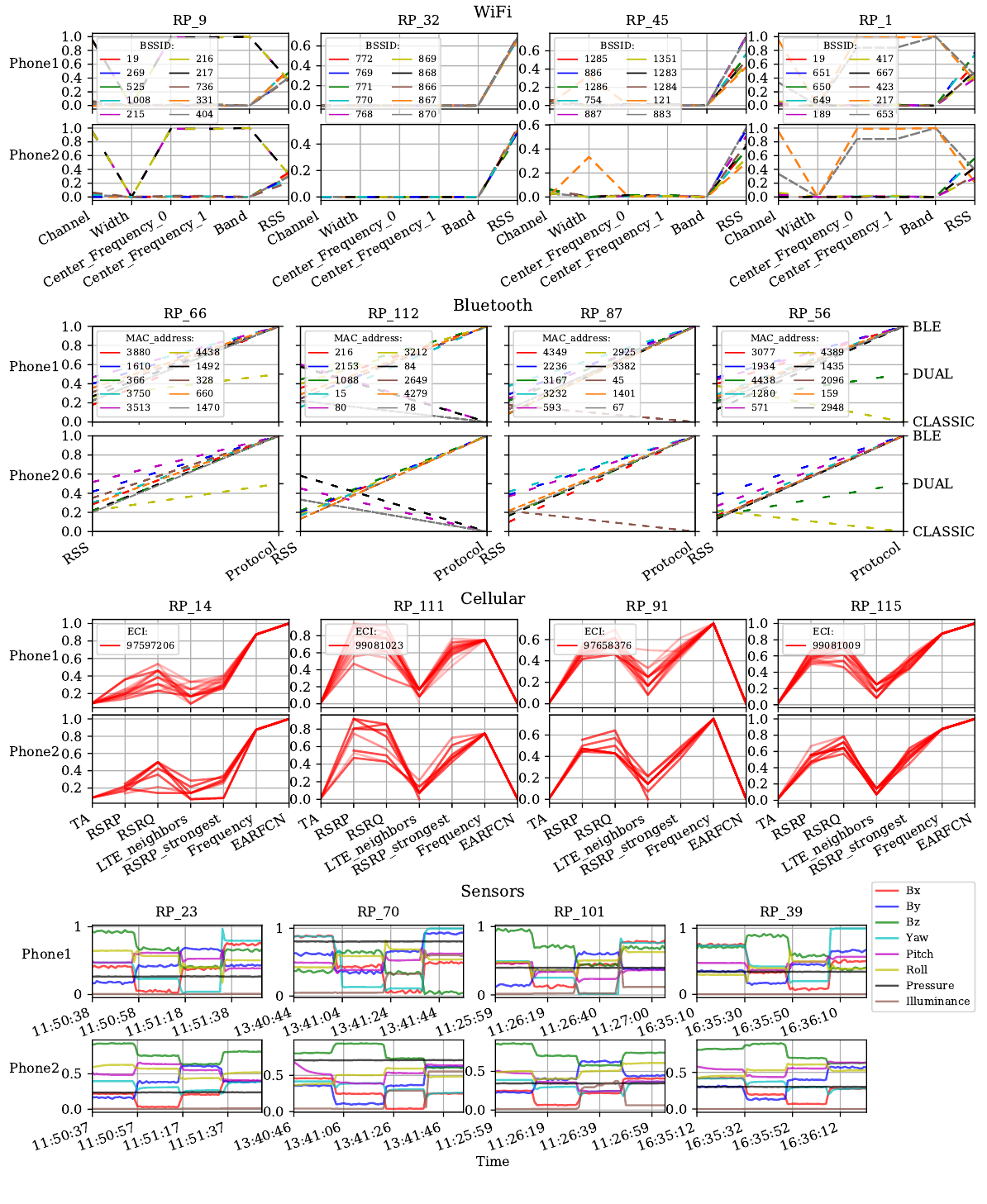}
\end{adjustbox}
\caption{Visualization of the data collected by Phone 1 and Phone 2 over randomly selected RPs. WiFi, Bluetooth, and cellular data are represented using parallel coordinate plots of the most important features, while sensor data are represented using time plots of magnetic field strength, angle of rotation, atmospheric pressure, and illuminance. All features are normalized between 0 and 1.}
\label{data_visualization}
\end{figure}

As graphical evidence of measurement validity, Fig. \ref{data_visualization} compares some of the data generated by the smartphones at randomly selected RPs side-by-side. Plots of the same data type exhibit the same profile despite corresponding to two different smartphones. Table \ref{table4} reports descriptive statistics of the data collected by each phone with respect to various variables. These statistics are compared against previously reported reference values, where applicable. The statistics displayed in Table \ref{table4} further support the validity of the dataset by ruling out the possibility that the dataset contains unrealistic, erratic, or random data.

\begin{table}[!h]
\begin{adjustbox}{width=1.2\textwidth,center=\textwidth}
\resizebox{\textwidth}{!}{
\centering
	\begin{tabular}{@{}rccccccccc@{}}
	\toprule
	& \multicolumn{4}{c}{Phone 1} & \multicolumn{4}{c}{Phone 2} \\
	\cmidrule(lr){2-5} \cmidrule(lr){6-9}
	& {Min} & {Max} & {Mean} & {SD} & {Min} & {Max} & {Mean} & {SD} & {Reference values}\\
	\cmidrule(lr){2-2} \cmidrule(lr){3-3} \cmidrule(lr){4-4} \cmidrule(lr){5-5} \cmidrule(lr){6-6} \cmidrule(lr){7-7} \cmidrule(lr){8-8} \cmidrule(lr){9-9} \cmidrule(lr){10-10}
	WiFi\\
	\cmidrule(lr){1-1}
	  \textit{Detected SSIDs} & 12 & 51 & 26.09 & 8.95 & 9 & 40 & 21.29 & 6.80 & {-} \\ 
	  \textit{Detected BSSIDs} & 98 & 223 & 159.32 & 31.68 & 67 & 168 & 114.97 & 23.92 & {-} \\
	  \textit{RSS} (dBm) & -97 & -53.33 & -85.82 & 6.86 & -99 & -38 & -84.20 & 6.88 & {$\approx [-102,-34]$ \cite{7275492}} \\ 
	\midrule
	Bluetooth\\
	\cmidrule(lr){1-1}
	  \textit{Detected MAC addresses} & 5 & 205 & 59.50 & 47.46 & 4 & 168 & 45.45 & 35.99 & {-} \\ 
	  \textit{RSS} (dBm) & -98 & -53 & -86.28 & 4.69 & -113 & -65 & -99.40 & 5.35 & {$\approx [-110,-48]$ \cite{baronti2018indoor}} \\ 
	\midrule
	Cellular\\ 
	\cmidrule(lr){1-1}
	  \textit{Detected ECIs}& 1 & 5 & 1.45 & 0.91 & 1 & 4 & 1.35 & 0.73 & {-}\\ 
	  \textit{LTE neighbors}& 0 & 12 & 2.36 & 1.53 & 0 & 14 & 2.45 & 1.79 & {-}\\
	  \textit{RSRP strongest} (dBm)& -128 & -81 & -103.32 & 6.90 & -127 & -82 & -105.18 & 8.26 & {-}\\
	  \textit{RSRP} (dBm)& -118 & -82 & -99.86 & 6.28 & -118 & -82 & -100.89 & 6.98 & {$\approx [-120,-70]$ \cite{6424050}}\\
	  \textit{RSRQ} (dB)& -20 & -7 & -12.83 & 2.33 & -20 & -6 & -12.87 & 2.48 & {$\approx [-24,-5]$ \cite{6424050}}\\
	\midrule
	Sensors\\
	\cmidrule(lr){1-1}
	  \textit{Magnitude of magnetic field} (\si{\micro\tesla}) & 38.52 & 51.07 & 44.49 & 3.51 & 29.45 & 73.03 & 51.90 & 13.40 & {$\approx$ 51} \cite{magcalc}\\ 
	  \textit{Atmospheric pressure} (\si{\hecto\pascal}) & 833.14 & 845.02 & 837.93 & 3.13 & 831.67 & 843.52 & 836.37 & 3.12 & {$\approx (829.66, 843.21, 836.43)$} \cite{WeathHist}\\ 
	  \textit{Illuminance} (\si{\micro\lux}) & \SI{1e-6}{} & 0.1508 & 0.0138 & 0.0271 & \SI{2e-7}{} & 0.1243 & 0.0104 & 0.0207 & {$\approx (0.1, 0.01, \si{1e-6})$ \cite{LightLevels}} \\ 
	\bottomrule  
\end{tabular}
}
\end{adjustbox}
\caption{Descriptive statistics of the OutFin dataset. These include the minimum, maximum, mean, and standard deviation of the most important variables. Reference values are provided where applicable. Small variations in results between the phones are mainly attributed to \textit{device heterogeneity} \cite{6663599} (e.g., the sensitivity of the radio receiver or sensor). The reference value for the magnitude of the magnetic field represents the Earth’s magnetic field around Denver, Colorado. The reference values for atmospheric pressure represent, respectively, the minimum, maximum, and mean recorded atmospheric pressure in Denver, Colorado, during the data collection period. The reference values for illuminance represent the light intensity for sunlight, daylight, and twilight, respectively. An hour-by-hour description of other weather conditions, such as temperature, humidity, and visibility at the time of data collection can be retrieved from \cite{othercondi}.}
\label{table4}
\end{table}

\subsection*{Usage Examples}
This subsection provides a brief demonstration of some of the application domains that OutFin can be used for. These include \textit{fingerprint interpolation}, \textit{feature extraction}, \textit{performance evaluation}, and \textit{signal denoising}.
\subsubsection*{Fingerprint Interpolation}
\begin{figure}[!t]
\centering
\includegraphics[width=\textwidth]{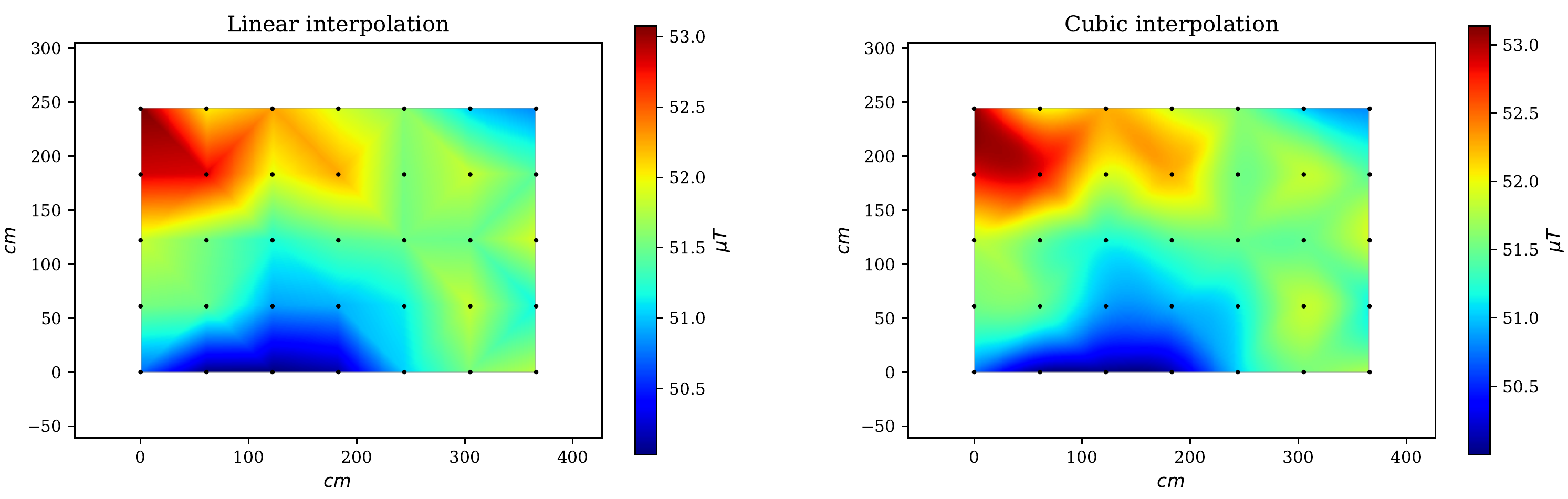}
\caption{Interpolated magnetic field magnitude of Site 3 using linear interpolation (left) and cubic interpolation (right). The maps were generated using calibrated magnetic field measurements from Phone 1 and Phone 2.}
\label{interpolation}
\end{figure}

Building a fingerprint map is usually required to provide positioning in a continuous fashion. The resolution of a map depends highly on the RP granularity (the higher the RP granularity, the better the map resolution). However, collecting fingerprints at highly granular RPs is time-consuming and labor intensive. Thus, interpolation methods are often employed to calculate the fingerprints between the locations of known fingerprints \cite{8373720}. The choice of an interpolation technique is pivotal to the resulting map. For example, Fig. \ref{interpolation} compares the magnetic field maps created for Site 3 by two different interpolation techniques, namely linear and cubic interpolation. Clearly, the resulting maps are not identical, which suggests that a positioning algorithm would exhibit a difference in performance depending on the employed map.

\subsubsection*{Feature Extraction}

\begin{figure}[!h]
\centering
\includegraphics[width=\textwidth]{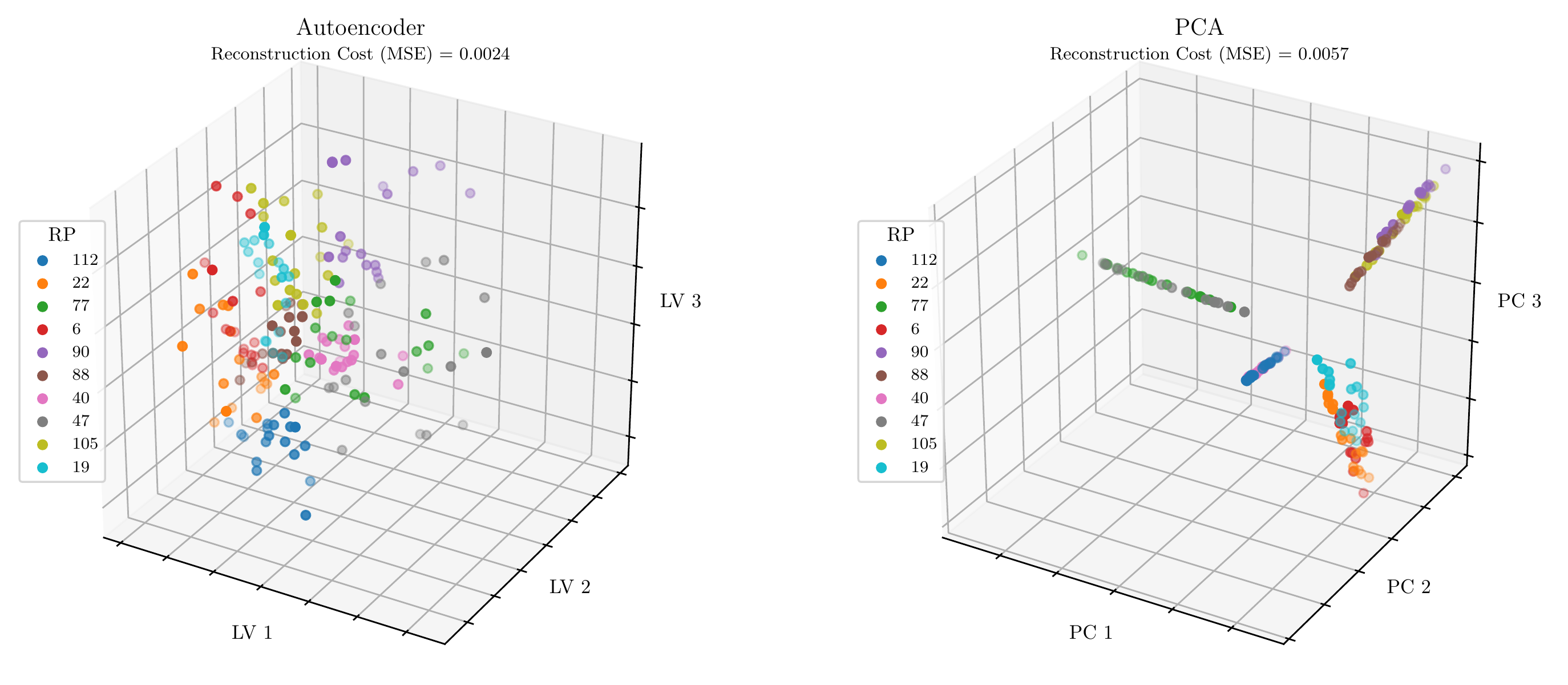}
\caption{The 3D codes for \num{18} WiFi RSS measurements (9 measurements per phone) for \num{10} randomly selected RPs produced by the autoencoder (left) and PCA (right). MSE: mean squared error; PC: principal component; LV: latent variable.}
\label{PCA_AE}
\end{figure}

A WiFi fingerprint has entries for all APs detected in an entire environment, but only a subset of these APs is observed at different locations. This is especially true for large-scale environments. For example, OutFin contains measurements from \si{1,379} unique APs; however, on average, only \SI{10}{\percent} of these APs are observed at any given RP. Consequently, feature extraction techniques are often utilized to reduce the dimensionality of the fingerprint space in order to achieve efficient and robust positioning \cite{10.1007/978-3-319-54042-9_57}. Fig. \ref{PCA_AE} compares two dimensionality reduction methods, i.e., the autoencoder and principal component analysis (PCA). The reconstruction cost obtained by the autoencoder is lower than that obtained by PCA. This suggests that the autoencoder is better at compressing the fingerprint space into a lower dimensional representation that comprises the informative content of the fingerprint space.

\subsubsection*{Performance Evaluation}
When proposing a new positioning method, the performance of the proposed method is often evaluated against the performance of previously proposed methods. It is often the case that at the heart of many of the methods  benchmarked against is a machine learning algorithm, such as $k$-Nearest Neighbors ($k$-NN), Support Vector Machine (SVM), Decision Tree, or Naive Bayes \cite{doi:10.1080/17489725.2020.1817582}. Therefore, with the purpose of comparing the performance of such algorithms, the positioning problem was casted as a classification task where each RP is treated as a unique class. Various performance metrics were considered, including classification metrics, positioning error, and computational complexity. For the sake of fair comparison, the parameters of each algorithm were fine-tuned using grid search and cross-validation. Evaluation results, shown in Table \ref{PE}, are reported on the Bluetooth measurements collected from Site 4. The results demonstrate that different algorithms can be ranked differently depending on the chosen performance metric. For example, the best classification accuracy was achieved by RBF SVM, while the lowest mean positioning error was achieved by $k$-NN.

\begin{table}[!h]
\begin{adjustbox}{width=1.15\textwidth,center=\textwidth}
\resizebox{\textwidth}{!}{
\centering
	\begin{tabular}{@{}rcccccccccc@{}}
	\toprule
	& \multicolumn{4}{c}{Classification Metric} & \multicolumn{4}{c}{Positioning Error (cm)} & \multicolumn{2}{c}{Computational Complexity \cite{ccml}}\\
	\cmidrule(lr){2-5} \cmidrule(lr){6-9} \cmidrule(lr){10-11}
	& {Accuracy} & {Precision} & {Recall} & {F1} & {Min} & {Max} & {Mean} & {SD} & {Training} & {Prediction}\\
	\cmidrule(lr){2-2} \cmidrule(lr){3-3} \cmidrule(lr){4-4} \cmidrule(lr){5-5} \cmidrule(lr){6-6} \cmidrule(lr){7-7} \cmidrule(lr){8-8} \cmidrule(lr){9-9} \cmidrule(lr){10-10} \cmidrule(lr){11-11}
	Algorithm\\
	\cmidrule(lr){1-1}
	  $k$-NN & 0.948 & 0.964 & 0.948 & 0.945 & 0.0 & 366.0 & 11.46 & 51.52 & {-} & {$\mathcal{O}(np)$} \\ 
	  RBF kernel SVM & 0.962 & 0.970 & 0.962 & 0.961 & 0.0 & 1098.0 & 18.81 & 121.46 & {$\mathcal{O}(n^2p+n^3)$} & {$\mathcal{O}(n_{sv}p)$} \\
	  Decision Tree & 0.957 & 0.967 & 0.957 & 0.956 & 0.0 & 732.0 & 15.19 & 83.19 & {$\mathcal{O}(n^2p)$} & {$\mathcal{O}(p)$} \\ 
	  Naive Bayes & 0.910 & 0.956 & 0.910 & 0.911 & 0.0 & 549.0 & 23.82 & 82.38 & {{$\mathcal{O}(np)$}} & {$\mathcal{O}(p)$} \\ 
	\bottomrule  
\end{tabular}
}
\end{adjustbox}
\caption{Performance evaluation of commonly used algorithms for positioning with respect to various metrics. The results were generated using 530 Bluetooth samples (\SI{60}{\percent} training and \SI{40}{\percent} testing) collected by both phones from Site 4. RBF: radial basis function; $n$: number of training samples;  $p$: number of features; $n_{sv}$: number of support vectors.}
\label{PE}
\end{table}

\subsubsection*{Signal Denoising}
\begin{figure}[!h]
\centering
\includegraphics[width=0.85\textwidth]{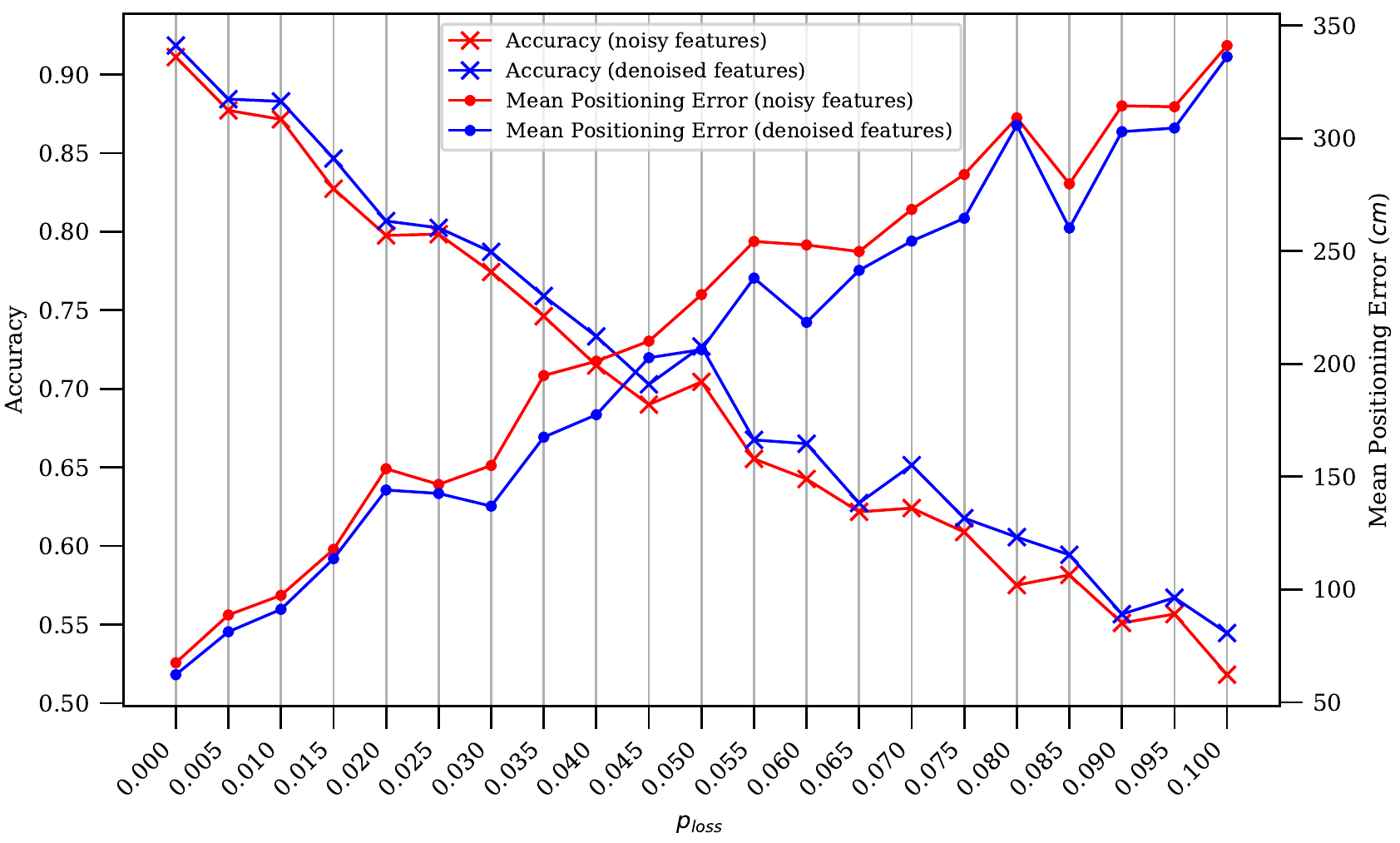}
\caption{Noisy vs. denoised features for positioning. For a given $p_{loss}$ value, the results were generated using \si{3,111} cellular samples collected by both phones from Site 2. A $k$-NN algorithm is used for comparison where $\sim$\SI{60}{\percent} of the samples were used for training and the remaining $\sim$\SI{40}{\percent} for testing.}
\label{NvsDN}
\end{figure}

Signal loss can negatively impact the performance of a positioning system. Thus, denoising techniques are often integrated as a preprocessing step to enhance positioning \cite{alhomayani2020deep}. As an example, a denoising autoencoder was utilized as a denoising agent where the feature vector of a cellular fingerprint is corrupted to emulate randomized loss of data. The degree of corruption is controlled by a predefined probability ($p_{loss}$) where, for example, a $p_{loss}$ of 0.03 indicates a \SI{3}{\percent} chance of setting a feature to zero. Fig. \ref{NvsDN} demonstrates the differences in performance between using noisy cellular features and their denoised versions for positioning in Site 2. On average, the use of the denoising step resulted in a \SI{1.43}{\percent} improvement in accuracy and a \SI{13.25}{\cm} reduction in positioning error.

\section*{Code availability}

Well-documented scripts, written in \texttt{Python 3.6.4} \cite{Python}, are present alongside the dataset (also available on GitHub \cite{OutFinCode}). These include the scripts used to generate the results described in the Technical Validation section as well as a script to calibrate magnetic field measurements against hard/soft-iron distortions. The data required to replicate the experiments reside in \texttt{OutFin/Code/temporal\_data}. Depending on the script, some of the following libraries may be required: \texttt{os, pandas, scipy, random, sklearn, matplotlib, numpy, statistics, keras, math}. Additionally, a thorough description of the collection environment in the form of an interactive map (developed using \texttt{QGIS 3.10} \cite{QGIS}) is provided. The map is composed of several layers that display information such as RP coordinates (both ground truth and smartphone estimated), pictures of the collection sites, and building height and ground elevation (as provided by the City and County of Denver \cite{denvergov}). High-resolution aerial imagery (3-inch), provided by the Denver Regional Council of Governments \cite{DRCOG}, are used as the basemap.

\clearpage
\printbibliography

\section*{Acknowledgments}
The authors would like to thank Dr. Steven Hick, the University of Denver Geographic Information Systems director, for his assistance with creating the interactive map; the University of Denver for allowing data collection on its campus; the City and County of Denver for providing building data; and the Denver Regional Council of Governments for providing the aerial imagery basemap.

\section*{Author contributions}
All authors equally contributed to all aspects of the presented work.

\section*{Competing interests}
The authors declare no competing interests.

\end{document}